\documentclass{article}
\usepackage{amsmath,amssymb,url,amsthm}
\usepackage[pdftex,hiresbb]{graphicx}
\allowdisplaybreaks
\def\theequation{\thesection.\arabic{equation}}

\def\d   {\displaystyle}
\newtheorem{thm}{Theorem}[section]

\newtheorem{lem}[thm]{Lemma}
\newtheorem{prop}[thm]{Proposition}

\newtheorem{ass}[thm]{Assumption}

\theoremstyle{definition}
\newtheorem{defn}[thm]{Definition}%[section]

\def\l     {\left}
\def\r     {\right}
\def\fin   {\hfill{$\Box$}\vspace{5mm}}

\def\vp    {\varphi}

\def\bbD   {{\mathbb D}}
\def\bbE   {{\mathbb E}}
\def\bbF   {{\mathbb F}}

\def\bbN   {{\mathbb N}}
\def\bbP   {{\mathbb P}}
\def\bbQ   {{\mathbb Q}}
\def\bbR   {{\mathbb R}}
\def\calB  {{\mathcal B}}
\def\calF  {{\mathcal F}}

\def\tN    {\widetilde{N}}

\def\olS     {\overline{S}}
\def\olsigma {\overline{\sigma}}
\def\tN    {\widetilde{N}}
\begin{document}
\title{Numerical analysis on locally risk-minimizing strategies for Barndorff-Nielsen and Shephard models}
\author{Takuji Arai \ \footnote{Department of Economics, Keio University, 2-15-45 Mita, Minato-ku, Tokyo, 108-8345, Japan. \\ (arai@econ.keio.ac.jp)}}
\maketitle

%%%%%%%%%%%%%%%%%%%%%%%%%%%%%%%%%%%%%%%%%%%%%%%%%%%%%%%%%%%%%%%%%%%%%%%%%%%%%%%
\begin{abstract}
We develop a numerical method for locally risk-minimizing (LRM) strategies for Barndorff-Nielsen and Shephard (BNS) models.
Arai et al. \cite{AIS-BNS} derived a mathematical expression for LRM strategies in BNS models using Malliavin calculus for L\'evy processes
and presented some numerical results only for the case where the asset price process is a martingale.
Subsequently, Arai and Imai \cite{AI} developed the first Monte Carlo (MC) method available for non-martingale BNS models with infinite active jumps.
Here, we modify the expression obtained by \cite{AIS-BNS} into a numerically tractable form, and, using the MC method developed by \cite{AI},
propose a numerical method of LRM strategies available for non-martingale BNS models with infinite active jumps.
In the final part of this paper, we will conduct some numerical experiments. 
\end{abstract}

%%%%%%%%%%%%%%%%%%%%%%%%%%%%%%%%%%%%%%%%%%%%%%%%%%%%%%%%%%%%%%%%%%%%%%%%%%%%%%%
%
% Section 1
%
%%%%%%%%%%%%%%%%%%%%%%%%%%%%%%%%%%%%%%%%%%%%%%%%%%%%%%%%%%%%%%%%%%%%%%%%%%%%%%%
\section{Introduction}\setcounter{equation}{0}
We aim to propose a numerical scheme for locally risk-minimizing (LRM) strategies for Barndorff-Nielsen and Shephard (BNS) models.
Local risk-minimization is a quadratic hedging method for financial derivatives in incomplete markets and has been actively studied for decades.
Arai and Suzuki \cite{AS} discussed LRM strategies for L\'evy markets using Malliavin calculus for L\'evy processes.
Here, L\'evy markets refer to financial models whose asset price process is given by a solution to the stochastic differential equation (SDE) driven by a Brownian motion and a pure jump L\'evy process.
In particular, they provided an explicit formula for LRM strategies, including Malliavin derivatives, under some mathematical assumptions.
As an extension of \cite{AS}, Arai et al. \cite{AIS-BNS} studied LRM strategies for BNS models: a representative framework of jump-type stochastic volatility models.
BNS models are mathematically of non-Gaussian Ornstein-Uhlenbeck (OU) type and have been actively studied in recent years,
following significant works by Barndorff-Nielsen and Shephard \cite{BNS1}, \cite{BNS2}.
Although some mild conditions are still needed, \cite{AIS-BNS} confirmed that BNS models satisfy all the conditions imposed in \cite{AS}
and derived a mathematical expression of LRM strategies for BNS models.

As for numerical techniques for BNS models, the Carr-Madan method, based on the Fast Fourier Transform, is well-known and very useful.
Still, it is available only when the discounted asset price process is a martingale. Indeed, numerical experiments implemented in \cite{AIS-BNS} treated only the martingale case.
Therefore, Arai and Imai \cite{AI} developed a Monte Carlo (MC) simulation method, which is also available for non-martingale BNS models.
Now, we discuss \cite{AI} in more detail. Depending on how jumps occur, BNS models are classified into two cases.
One is finite active models, in which only a finite number of jumps occur in any finite time period, and the other is infinite active models, in which infinitely many jumps occur.
Gamma-OU and IG-OU types are typical examples of finite and infinite active models, respectively.
Since it is relatively easy to develop an MC method for finite active models, \cite{AI} focused only on IG-OU type.
The MC method in \cite{AI} is based on the exact simulation method for OU processes associated with tempered stable distributions developed by Sabino and Petroni \cite{SP}.
Furthermore, for the case where the discounted asset price process is not a martingale, we need to simulate the model under an equivalent martingale measure.
Indeed, \cite{AI} treated the minimal martingale measure (MMM), which is closely related to LRM strategies.

In this paper, using the MC method in \cite{AI}, we develop a numerical scheme for LRM strategies in non-martingale IG-OU type BNS models and conduct some numerical experiments.
Since the expression of LRM strategies obtained in \cite{AIS-BNS} contains some Malliavin derivatives, we shall modify it into a numerically tractable form.
Specifically, we shall convert terms involving Malliavin derivatives to conditional expectations under the MMM, which makes the MC method in \cite{AI} available.
However, the conditional expectations appear as an integrand of an integral with respect to the L\'evy measure of IG-OU type.
Therefore, we need to approximate the integral as accurately as possible.

This paper is organized as follows: Section 2 introduces BNS models, the MMM, and the assumptions of this paper.
We define LRM strategies in Subsection 3.1 and convert the mathematical expression obtained by \cite{AIS-BNS} into a form that enables us to develop a numerical scheme in Subsection 3.2.
Note that the definitions and properties related to Malliavin calculus are postponed until Appendix A.
Section 4 is devoted to numerical experiments, and Section 5 concludes this paper.

%%%%%%%%%%%%%%%%%%%%%%%%%%%%%%%%%%%%%%%%%%%%%%%%%%%%%%%%%%%%%%%%%%%%%%%%%%%%%%%
%
% Section 2
%
%%%%%%%%%%%%%%%%%%%%%%%%%%%%%%%%%%%%%%%%%%%%%%%%%%%%%%%%%%%%%%%%%%%%%%%%%%%%%%%
\section{BNS models}\setcounter{equation}{0}
We consider a financial market composed of one riskless asset and one risky asset with maturity $T>0$.
For simplicity, the interest rate of our market is assumed to be zero.
In BNS models, the risky asset price process denoted by $S=\{S_t\}_{0\leq t\leq T}$ is represented as follows:
\[
S_t:=S_0\exp\l\{\int_0^t\l(\mu-\frac{1}{2}\sigma_s^2\r)ds+\int_0^t\sigma_sdW_s+\rho H_{\lambda t}\r\}, \ \ \ t\in[0,T],
\]
where $S_0>0$, $\mu\in\bbR$, $\rho\leq0$, $\lambda>0$, $W=\{W_t\}_{0\leq t\leq T}$ is a one-dimensional standard Brownian motion, and
$H_\lambda=\{H_{\lambda t}\}_{0\leq t\leq T}$ is a driftless subordinator, in other words, a non-decreasing L\'evy process without drift.
In addition, $\sigma=\{\sigma_t\}_{0\leq t\leq T}$ is the volatility process, which is defined as the square root of the solution $\sigma^2=\{\sigma^2_t\}_{0\leq t\leq T}$ to the following SDE: 
\[
d\sigma_t^2 = -\lambda\sigma_t^2dt+dH_{\lambda t}, \ \ \ \sigma_0^2>0.
\]
Remark that the underlying probability space $(\Omega, \calF, \bbP)$ is given by the product space (\ref{eq-A1}) introduced in Appendix A.
Hence, $W$ is the coordinate mapping process of the one-dimensional Wiener space $(\Omega_W, \calF_W, \bbP_W)$, and $H_\lambda$ is defined as $H_{\lambda t}:=J_t$,
where $J$ is the subordinator associated with the canonical L\'evy space $(\Omega_J, \calF_J, \bbP_J)$.
In addition, $\{\calF_t\}_{0\leq t\leq T}$ denotes the canonical filtration completed for $\bbP$.

Now, let $N$ be the Poisson random measure of $H_\lambda$ defined on $[0,T]\times(0,\infty)$. In addition, $\nu$ denotes its L\'evy measure.
Note that $\nu$ satisfies
\[
\int_0^\infty(x\wedge1)\nu(dx)<\infty,
\]
since $H_\lambda$ is a subordinator. We can see then that $S$ is the solution to the following SDE:
\[
dS_t = S_{t-}\l\{\alpha dt+\sigma_t dW_t + \int_0^\infty(e^{\rho x}-1)\tN(dt,dx)\r\},
\]
where
\[
\alpha:=\mu+\int_0^\infty(e^{\rho x}-1)\nu(dx),
\]
and $\tN(dt,dx)$ denotes the compensated Poisson random measure of $H_\lambda$, that is,
\[
\tN(dt,dx):=N(dt,dx)-\nu(dx)dt.
\]
For more details on BNS models, see e.g. Arai et al. \cite{AIS-BNS}, Nicolato and Venardos \cite{NV}, and Schoutens \cite{Scho}.

As mentioned in the Introduction, BNS models have two representative types.
The first is the case in which the L\'evy measure is given by
\[
\nu(dx)=abe^{-bx}{\bf 1}_{(0,\infty)}(x)dx,
\]
where $a>0$ and $b>0$. This case is called Gamma-OU type, an example of finite active models.
The invariant distribution for the squared volatility process $\sigma^2$ follows a Gamma distribution with parameters $a>0$ and $b>0$.
The other type is IG-OU, in which the invariant distribution of $\sigma^2$ is given by an inverse-Gaussian distribution with parameters $a>0$ and $b>0$.
The corresponding L\'evy measure is described as
\begin{equation}\label{eq-nu-IGOU}
\nu(dx)=\frac{a\lambda}{2\sqrt{2\pi}}x^{-\frac{3}{2}}(1+b^2x)e^{-\frac{1}{2}b^2x}{\bf 1}_{(0,\infty)}(x)dx.
\end{equation}
This case has infinite active jumps, since $\nu((0,\infty))=\infty$.
Throughout this paper, we consider IG-OU type BNS models. For later use, we define the L\'evy density $f_\nu$ and a function $g_\nu$ as
\begin{equation}\label{eq-gnu}
\nu(dx)=f_\nu(x)dx, \ \ \ \mbox{and} \ \ \ g_\nu(x):=(e^{\rho x}-1)f_\nu(x), \ \ \ x>0,
\end{equation}
where $\nu$ is defined in (\ref{eq-nu-IGOU}).

Next, we discuss the MMM. To begin with, we introduce the following assumption:

%%%%%%%%%%%%%%%%%%%%%%%%%%%%%%%%%%%%%%%%%%%%%%%%%%%%%%%%%%%%%%%%%%%%%%%%%%%%%%%
\begin{ass}\label{ass}
Throughout this paper, we assume that
\[
\frac{b^2}{2}>2\l(\frac{1-e^{-\lambda T}}{\lambda}\vee|\rho|\r) \ \ \ \mbox{and} \ \ \ \frac{\alpha}{e^{-\lambda T}\sigma^2_0+C^\rho_2}>-1,
\]
where
\begin{equation}\label{eq-Crho2}
C^\rho_2:=\int_0^\infty(e^{\rho x}-1)^2f_\nu(x)dx=2\rho\lambda a\l(\frac{1}{\sqrt{b^2-4\rho}}-\frac{1}{\sqrt{b^2-2\rho}}\r).
\end{equation}
\end{ass}

\noindent
Let $M=\{M_t\}_{0\leq t\leq T}$ be the martingale part of $S$. That is, it is defined as
\[
M_t=\int_0^tS_{s-}\sigma_s dW_s + \int_0^tS_{s-}\int_0^\infty(e^{\rho x}-1)\tN(ds,dx), \ \ \ M_0=0.
\]
An equivalent martingale measure $\bbQ\sim\bbP$ is called the minimal martingale measure (MMM) if any square integrable $\bbP$-martingale orthogonal to $M$ remains a martingale under $\bbQ$.
Now, we define a process $Z=\{Z_t\}_{0\leq t\leq T}$ as
\[
Z_t:=\bbE\l[\frac{d\bbQ}{d\bbP}\big|\calF_t\r],
\]
which is a martingale with $Z_0=1$ and $Z_T=\dfrac{d\bbQ}{d\bbP}$. We call $Z$ the density process of $\bbQ$. Note that $Z$ is also the solution to the following SDE:
\begin{equation}\label{SDE-Z}
dZ_t=-\frac{\alpha Z_{t-}}{S_{t-}(\sigma_t^2+C^\rho_2)}dM_t.
\end{equation}
Assumption \ref{ass} ensures the positivity of the solution to the SDE (\ref{SDE-Z}).
Thus, as seen in Proposition 2.7 of \cite{AIS-BNS}, the MMM $\bbQ$ exists as a probability measure and $S$ is a $\bbQ$-martingale.
Moreover, (2.5) of \cite{AIS-BNS} implies that $Z_t$ is expressed as
\begin{align}\label{eq-Z}
Z_t &= \exp\bigg\{-\int_0^tu_sdW_s-\frac{1}{2}\int_0^tu_s^2ds+\int_0^t\int_0^\infty\log(1-\theta_{s,x})\tN(ds,dx) \nonumber \\
    &  \hspace{5mm}+\int_0^t\int_0^\infty(\log(1-\theta_{s,x})+\theta_{s,x})f_\nu(x)dxds\bigg\},
\end{align}
where
\[
u_t:=\frac{\alpha\sigma_t}{\sigma_t^2+C^\rho_2} \ \ \ \mbox{ and } \ \ \ \theta_{t,x}:=\frac{\alpha(e^{\rho x}-1)}{\sigma_t^2+C^\rho_2}.
\]

%%%%%%%%%%%%%%%%%%%%%%%%%%%%%%%%%%%%%%%%%%%%%%%%%%%%%%%%%%%%%%%%%%%%%%%%%%%%%%%
%
% Section 3
%
%%%%%%%%%%%%%%%%%%%%%%%%%%%%%%%%%%%%%%%%%%%%%%%%%%%%%%%%%%%%%%%%%%%%%%%%%%%%%%%
\section{LRM strategies}\setcounter{equation}{0}
\subsection{Definition}
%%%%%%%%%%%%%%%%%%%%%%%%%%%%%%%%%%%%%%%%%%%%%%%%%%%%%%%%%%%%%%%%%%%%%%%%%%%%%%%
We define LRM strategies based on Theorem 1.6 of Schweizer \cite{Sch3}.
Recall that $M$ is the martingale part of the asset price process $S$ with $M_0=0$. Moreover, we define a process $A$ as $A_t:=S_t-S_0-M_t$.

%%%%%%%%%%%%%%%%%%%%%%%%%%%%%%%%%%%%%%%%%%%%%%%%%%%%%%%%%%%%%%%%%%%%%%%%%%%%%%%
\begin{defn}\label{def-1}
\begin{enumerate}
\item $\Theta_S$ denotes the space of all $\bbR$-valued predictable processes $\xi$ satisfying
      \[
      \bbE\l[\int_0^T\xi_t^2d\langle M\rangle_t+\l(\int_0^T|\xi_tdA_t|\r)^2\r]<\infty.
      \]
\item An $L^2$-strategy is given by a pair $\vp=(\xi,\eta)$, where $\xi\in\Theta_S$ and $\eta$ is an adapted process such that
      $V(\vp):=\xi S+\eta$ is a right continuous process with $\bbE[V_t^2(\vp)]<\infty$ for every $t\in[0,T]$.
      Note that $\xi_t$ (resp. $\eta_t$) represents the number of units of the risky asset (resp., the riskless asset) an investor holds at time $t$.
\item For given financial derivative $X\in L^2(\bbP)$, the process $C^X(\vp)$ defined by
      \[
      C^X_t(\vp):=X1_{\{t=T\}}+V_t(\vp)-\int_0^t\xi_s dS_s
      \]
      is called the cost process of $\vp=(\xi, \eta)$ for $X$.
\item An $L^2$-strategy $\vp$ is said to be the locally risk-minimizing (LRM) strategy for $X$ if $V_T(\vp)=0$ and $C^F(\vp)$ is a martingale orthogonal to $M$, that is,
      $[C^X(\vp),M]$ is a uniformly integrable martingale.
\item An $X\in L^2(\bbP)$ admits a F\"ollmer--Schweizer (FS) decomposition if it can be described by
      \begin{equation} \label{eqFS}
      X=X_0+\int_0^T\xi_t^XdS_t+L_T^X,
      \end{equation}
      where $X_0\in\bbR$, $\xi^X\in\Theta_S$ and $L^X$ is a square-integrable martingale orthogonal to $M$ with $L_0^X=0$.
\end{enumerate}
\end{defn}

\noindent
Proposition 5.2 of \cite{Sch3} showed the following result:
Under Assumption \ref{ass}, the LRM strategy $\vp=(\xi,\eta)$ for $X$ exists if and only if $X$ admits an FS decomposition, and its relationship is given by
\[
\xi_t=\xi^X_t,\hspace{3mm}\eta_t=X_0+\int_0^t\xi^X_sdS_s+L^X_t-X1_{\{t=T\}}-\xi^X_tS_t.
\]
Therefore, it suffices to derive a representation of $\xi^X$ in (\ref{eqFS}) to obtain the LRM strategy for the financial derivative $X$.
Henceforth, we identify $\xi^X$ with the LRM strategy for $X$. For more details on LRM strategies, see \cite{Sch3}.

%%%%%%%%%%%%%%%%%%%%%%%%%%%%%%%%%%%%%%%%%%%%%%%%%%%%%%%%%%%%%%%%%%%%%%%%%%%%%%%
\subsection{Mathematical expression}
%%%%%%%%%%%%%%%%%%%%%%%%%%%%%%%%%%%%%%%%%%%%%%%%%%%%%%%%%%%%%%%%%%%%%%%%%%%%%%%
In this subsection, we derive an expression for LRM strategies in BNS models that is tractable for developing a numerical scheme.
As a financial derivative to be hedged, we treat a put option with strike price $K$, because its payoff is bounded, and therefore, we do not need to worry about any integrability conditions.
Furthermore, the LRM strategy for a call option can be derived from the one for the put option with the same strike price.

Recall that $\xi_t^{(K-S_T)^+}$ denotes the LRM strategy at time $t\in[0,T]$ for the put option $(K-S_T)^+$.
The following mathematical expression of $\xi_t^{(K-S_T)^+}$ has been derived under Assumption \ref{ass} by \cite{AIS-BNS} in their Theorem 3.1:
\begin{align}
\xi_t^{(K-S_T)^+}
&= \frac{1}{S_{t-}(\sigma_t^2+C^\rho_2)}\bigg\{\sigma_t^2\bbE_\bbQ\Big[-{\bf 1}_{\{S_T<K\}}S_T|\calF_{t-}\Big] \nonumber \\
&  \hspace{5mm}+\int_0^\infty\bbE_\bbQ\Big[(K-S_T)^+(G^\bbQ_{t,z}-1) \nonumber \\
&  \hspace{5mm}+zG^\bbQ_{t,z}D_{t,z}(K-S_T)^+|\calF_{t-}\Big]g_\nu(z)dz\bigg\},
\label{eq-thm-main}
\end{align}
where $C^\rho_2$ and $g_\nu$ respectively, are defined in (\ref{eq-Crho2}) and (\ref{eq-gnu}), $D_{t,z}(K-S_T)^+$ is given by Proposition \ref{prop-put-deriv}, and
\[
G^\bbQ_{t,z}:= \exp\l\{zD_{t,z}\log Z_T-\log(1-\theta_{t,z})\r\}, \ \ \ (t,z)\in[0,T]\times(0,\infty).
\]
Here, for $D_{t,z}\log Z_T$, refer to Proposition \ref{prop-logZT}.

Henceforth, we fix $t\in[0,T]$ arbitrarily, and calculate the value of $\xi_t^{(K-S_T)^+}$ under the conditions $S_t=\olS$ and $\sigma^2_t=\olsigma^2$.
Taking into account
\[
\bbQ(S_t\neq S_{t-})=\bbQ(\sigma^2_t\neq\sigma^2_{t-})=\bbQ(\Delta H_{\lambda t}>0)=0,
\]
we can rewrite (\ref{eq-thm-main}) as
\begin{align}
\xi_t^{(K-S_T)^+}
&= \frac{1}{\olS(\olsigma^2+C^\rho_2)}\bigg\{\olsigma^2\bbE_\bbQ\Big[-{\bf 1}_{\{S_T<K\}}S_T|S_t=\olS, \sigma_t^2=\olsigma^2\Big] \nonumber \\
&  \hspace{5mm}+\int_0^\infty\bbE_\bbQ\Big[(K-S_T)^+(G^\bbQ_{t,z}-1) \nonumber \\
&  \hspace{5mm}+zG^\bbQ_{t,z}D_{t,z}(K-S_T)^+|S_t=\olS, \sigma_t^2=\olsigma^2\Big]g_\nu(z)dz\bigg\}.
\label{eq-thm-main2}
\end{align}
Therefore, restricting the time period to $[t,T]$, we consider the asset price and squared volatility processes under the initial conditions $S_t=\olS$ and $\sigma^2_t=\olsigma^2$ at time $t$.
Thus, for $s\in[t,T]$, $S_s$ and $\sigma^2_s$ are represented as
\begin{align*}
S_s        &= \olS\exp\l\{\int_t^s\l(\mu-\frac{1}{2}\sigma_u^2\r)du+\int_t^s\sigma_udW_u+\rho(H_{\lambda s}-H_{\lambda t})\r\}, \\
\sigma^2_s &= e^{-\lambda(s-t)}\olsigma^2+\int_t^se^{-\lambda(u-t)}dH_{\lambda u},
\end{align*}
respectively.

Now, we denote 
\[
F_t(\olS,\olsigma^2):=\bbE_\bbQ[(K-S_T)^+|S_t=\olS, \sigma_t^2=\olsigma^2],
\]
which gives the put option price at time $t$ with strike price $K$ under the MMM under the conditions $S_t=\olS$ and $\sigma_t^2=\olsigma^2$,
and which is numerically computable by using the MC method of \cite{AI}.
Furthermore, for $z>0$, we convert the initial values at time $t$ from $\olS$ and $\olsigma^2$ into $\olS e^{\rho z}$ and $\olsigma^2+z$, respectively,
and consider the function $F_t(\olS e^{\rho z},\olsigma^2+z)$.
Let $S^{t,z}=\{S^{t,z}_s\}_{t\leq s\leq T}$ and $(\sigma^{t,z})^2=\{(\sigma^{t,z}_s)^2\}_{t\leq s\leq T}$ be the asset price and squared volatility processes, respectively,
under the initial conditions $S^{t,z}_t=\olS e^{\rho z}$ and $(\sigma^{t,z}_t)^2=\olsigma^2+z$. In other words, $(\sigma^{t,z})^2$ is defined as
\begin{equation}\label{eq-sigmatz}
(\sigma^{t,z}_s)^2:=\sigma_s^2+ze^{-\lambda(s-t)}=e^{-\lambda(s-t)}(\olsigma^2+z)+\int_t^se^{-\lambda(u-t)}dH_{\lambda u}, \ \ \ s\in[t,T].
\end{equation}
In addition, $S^{t,z}$ can also be defined as the solution to the following SDE:
\begin{equation}\label{eq-Stz}
dS^{t,z}_s = S^{t,z}_{s-}\l\{\alpha dt+\sigma^{t,z}_sdW_s+\int_0^\infty(e^{\rho x}-1)\tN(ds,dx)\r\}, \ \ \ S^{t,z}_t=\olS e^{\rho z}.
\end{equation}
Note that when we change the initial conditions at $t$, we must also modify the MMM.
The corresponding MMM, denoted by $\bbQ^{t,z}$, can be described as
\begin{align}
\frac{d\bbQ^{t,z}}{d\bbP}
&= \exp\bigg\{-\int_t^Tu^{t,z}_sdW_s-\frac{1}{2}\int_t^T(u^{t,z}_s)^2ds+\int_t^T\int_0^\infty\log(1-\theta^{t,z}_{s,x})\tN(ds,dx) \nonumber \\
&  \hspace{5mm}+\int_t^T\int_0^\infty(\log(1-\theta^{t,z}_{s,x})+\theta^{t,z}_{s,x})f_\nu(x)dxds\bigg\},
\label{eq-Qtz}
\end{align}
where
\begin{equation}\label{eq-utz}
u^{t,z}_s=\frac{\alpha\sigma^{t,z}_s}{(\sigma^{t,z}_s)^2+C^\rho_2} \ \ \ \mbox{ and } \ \ \
\theta^{t,z}_{s,x}=\frac{\alpha(e^{\rho x}-1)}{(\sigma^{t,z}_s)^2+C^\rho_2}, \ \ \ s\in[t,T], \ x\in(0,\infty).
\end{equation}
Using the above notations, we have
\begin{equation}\label{eq-Ftz}
F_t(\olS e^{\rho z},\olsigma^2+z)=\bbE_{\bbQ^{t,z}}[(K-S^{t,z}_T)^+|S^{t,z}_t=\olS e^{\rho z}, (\sigma^{t,z}_t)^2=\olsigma^2+z], \ \ \ z>0.
\end{equation}

As seen below, we can rewrite the expression (\ref{eq-thm-main2})  in a form that contains an integral of the function $F_t$ with respect to the L\'evy measure.

%%%%%%%%%%%%%%%%%%%%%%%%%%%%%%%%%%%%%%%%%%%%%%%%%%%%%%%%%%%%%%%%%%%%%%%%%%%%%%%
\begin{thm}\label{thm-1}
For any $t\in[0,T]$ and $K>0$, given $S_t=\olS$ and $\sigma_t^2=\olsigma^2$, we have
\begin{align}\label{eq-thm-1}
\xi_t^{(K-S_T)^+} &= \frac{1}{\olS(\olsigma^2+C^\rho_2)}\bigg\{\olsigma^2\bbE_\bbQ\Big[-{\bf 1}_{\{S_T<K\}}S_T|S_t=\olS, \sigma_t^2=\olsigma^2\Big] \nonumber \\
                  &   \hspace{5mm}+\int_0^\infty F_t(\olS e^{\rho z},\olsigma^2+z)g_\nu(z)dz-C^\rho_1F_t(\olS,\olsigma^2)\bigg\},
\end{align}
where $\d{C^\rho_1:=\int_0^\infty g_\nu(z)dz=\frac{\rho\lambda a}{\sqrt{b^2-2\rho}}}$.
\end{thm}

\proof
Proposition \ref{prop-put-deriv} implies that 
\begin{align*}
\lefteqn{\bbE_\bbQ\l[(K-S_T)^+(G^\bbQ_{t,z}-1)+zG^\bbQ_{t,z}D_{t,z}(K-S_T)^+|S_t=\olS, \sigma_t=\olsigma\r]} \\
&= \bbE_\bbQ\l[G^\bbQ_{t,z}\Big((K-S_T)^++zD_{t,z}(K-S_T)^+\Big)\Big|S_t=\olS, \sigma_t=\olsigma\r] \\
&\hspace{5mm}-\bbE_\bbQ[(K-S_T)^+|S_t=\olS, \sigma_t=\olsigma] \\
&= \bbE\l[\frac{Z_TG^\bbQ_{t,z}}{Z_t}(K-S_T\exp\{zD_{t,z}\log S_T\})^+\Big|S_t=\olS, \sigma_t=\olsigma\r]-F_t(\olS,\olsigma^2).
\end{align*}
From the view of (\ref{eq-thm-main2}), we have only to show that 
\[
F_t(\olS e^{\rho z},\olsigma^2+z)=\bbE\l[\frac{Z_TG^\bbQ_{t,z}}{Z_t}(K-S_T\exp\{zD_{t,z}\log S_T\})\Big|S_t=\olS, \sigma_t=\olsigma\r]
\]
for any $z>0$.

Henceforth, we fix $z\in(0,\infty)$ arbitrarily. From the view of Proposition \ref{prop-logZT}, we have
\begin{align}
G^\bbQ_{t,z}
&=  \exp\l\{zD_{t,z}\log Z_T-\log(1-\theta_{t,z})\r\} \nonumber \\
&=  \exp\bigg\{-\int_t^TzD_{t,z}u_sdW_s-\int_t^Tzu_sD_{t,z}u_sds-\frac{1}{2}\int_t^T(zD_{t,z}u_s)^2ds \nonumber \\
&   \hspace{5mm}+\int_t^T\int_0^\infty zD_{t,z}\log(1-\theta_{s,x})\tN(ds,dx) \nonumber \\
&   \hspace{5mm}+\int_t^T\int_0^\infty(zD_{t,z}\log(1-\theta_{s,x})+zD_{t,z}\theta_{s,x})f_\nu(x)dxds\bigg\}.
\label{eq-thm-1-2}
\end{align}
As for $u^{t,z}_s$ and $\theta^{t,z}_{s,x}$ defined in (\ref{eq-utz}), we have
\[
\l\{\begin{array}{l}
u^{t,z}_s          = u_s+zD_{t,z}u_s, \\
\theta^{t,z}_{s,x} = \theta_{s,x}+zD_{t,z}\theta_{s,x} \\
\end{array}\r.
\]
for any $s\in[t,T]$ and $x\in(0,\infty)$ by Lemmas \ref{lem-ut-2} and \ref{lem-ut-3}. Furthermore, Lemma \ref{lem-ut-4} implies that
\[
\log(1-\theta_{s,x})+zD_{t,z}\log(1-\theta_{s,x})=\log(1-\theta^{t,z}_{s,x}).
\]
As a result, (\ref{eq-Z}), (\ref{eq-thm-1-2}) and (\ref{eq-Qtz}) provide that
\begin{align*}
\frac{Z_TG^\bbQ_{t,z}}{Z_t}
&= \exp\bigg\{-\int_t^T(u_s+zD_{t,z}u_s)dW_s-\frac{1}{2}\int_t^T(u_s+zD_{t,z}u_s)^2ds \\
&  \hspace{5mm}+\int_t^T\int_0^\infty[\log(1-\theta_{s,x})+zD_{t,z}\log(1-\theta_{s,x})]\tN(ds,dx) \\
&  \hspace{5mm}+\int_t^T\int_0^\infty[\log(1-\theta_{s,x})+zD_{t,z}\log(1-\theta_{s,x}) \\
&  \hspace{5mm}+\theta_{s,x}+zD_{t,z}\theta_{s,x}]f_\nu(x)dxds\bigg\} \\
&= \exp\bigg\{-\int_t^Tu^{t,z}_sdW_s-\frac{1}{2}\int_t^T(u^{t,z}_s)^2ds \\
&  \hspace{5mm}+\int_t^T\int_0^\infty\log(1-\theta^{t,z}_{s,x})\tN(ds,dx) \\
&  \hspace{5mm}+\int_t^T\int_0^\infty[\log(1-\theta^{t,z}_{s,x})+\theta^{t,z}_{s,x}]f_\nu(x)dxds\bigg\} \\
&= \frac{d\bbQ^{t,z}}{d\bbP}.
\end{align*}
Note that we have
\[
S^{t,z}_T=S_T\exp\l\{zD_{t,z}\log S_T\r\}
\]
by Proposition \ref{prop-LT-1}, where the process $S^{t,z}$ is defined in (\ref{eq-Stz}).
Then, (\ref{eq-Ftz}) implies that
\begin{align*}
\lefteqn{\bbE\l[\frac{Z_TG^\bbQ_{t,z}}{Z_t}(K-S_T\exp\{zD_{t,z}\log S_T\})^+\Big|S_t=\olS, \sigma_t=\olsigma\r]} \\
&= \bbE_{\bbQ^{t,z}}[\l(K-S^{t,z}_T\r)^+|S^{t,z}_t=\olS e^{\rho z}, (\sigma^{t,z}_t)^2=\olsigma^2+z] \\
&= F_t(\olS e^{\rho z},\olsigma^2+z),
\end{align*}
where the process $(\sigma^{t,z})^2$ is defined in (\ref{eq-sigmatz}). This comletes the proof of Theorem \ref{thm-1}.
\fin

%%%%%%%%%%%%%%%%%%%%%%%%%%%%%%%%%%%%%%%%%%%%%%%%%%%%%%%%%%%%%%%%%%%%%%%%%%%%%%%
%
% Section 4
%
%%%%%%%%%%%%%%%%%%%%%%%%%%%%%%%%%%%%%%%%%%%%%%%%%%%%%%%%%%%%%%%%%%%%%%%%%%%%%%%
%\newpage
\setcounter{equation}{0}\section{Numerical results}
We present the results of numerical experiments in this section. We use two model parameter sets shown in Table \ref{t-1}.
NV and Scho, respectively, are the parameter sets introduced in Table 5.1 of \cite{NV} and a modified version of the set introduced in Table 7.1 of \cite{Scho}.
Note that NV satisfies Assumption \ref{ass}. On the other hand, the value of $b$ in Table 7.1 of \cite{Scho} is calibrated to 0.7995, but it violates Assumption \ref{ass}.
Thus, changing it to 4.7995 in Scho, we conduct our experiments. 
In addition, since the accuracy of the MC method deteriorates as the value of $\alpha$ increases, as mentioned in \cite{AI}, we set $\alpha$ appropriately within a not-too-large range.
All numerical experiments use MATLAB R2024a.

%%%%%%%%%%%%%%%%%%%%%%%%%%%%%%%%%%%%%%%%%%%%%%%%%%%%%%%%%%%%%%%%%%%%%%%%%%%%%%%
\begin{table}[h]\centering\caption{Parameter sets}\begin{tabular}{cccccccc}\hline \vspace{-3.5mm} \\
     & $\alpha$ & $\olS$   & $\olsigma^2$ & $\rho$    & $\lambda$ & $a$    & $b$      \\ \hline \hline
NV   & 0.007    & \ 468.40 & 0.0041       & $-$4.7039 & 2.4958    & 0.0872 & 11.9800  \\ \hline
Scho & 0.100    & 1124.47  & 0.0156       & $-$0.1926 & 0.0636    & 6.2410 & \ 4.7995 \\ \hline
\end{tabular}\label{t-1}\end{table}

To conduct our numerical experiments, we need to compute the right-hand side of (\ref{eq-thm-1}). 
First, we simulate the value of $F_t(\olS e^{\rho z},\olsigma^2+z)$ for given $z>0$ by using the MC method developed by \cite{AI}.
Using this, we compute the integration in (\ref{eq-thm-1}) using a discrete approximation based on the trapezoidal rule with $N$ grid points $0<z_1<z_2<\dots<z_N$.
As for $N$ and $\{z_n\}_{1\leq n\leq N}$, we shall set them so that the integral $\d{\int_0^\infty g_\nu(z)dz}$($=:C^\rho_1$) can be approximated with sufficient accuracy.
More precisely, taking into account that $\d{\lim_{z\to0}g_\nu(z)=\infty}$, we approximate $C^\rho_1$ based on the trapezoidal rule by the following value:
\begin{equation}\label{eq-approxcrho1}
\int_0^{z_1}g_\nu(z)dz+\frac{1}{2}\sum_{n=1}^{N-1}(z_{n+1}-z_n)[g_\nu(z_n)+g_\nu(z_{n+1})].
\end{equation}
Note that the first term above is calculated using the numerical integration function of MATLAB. 
Indeed, for the NV parameter set, we set $N=400$ and the sequence $\{z_n\}_{n=1,\dots,N}$ to
\[
z_n=\l\{\begin{array}{ll}
    n\times 10^{-5},        & n=1,  \dots,200, \\
    z_{200}+(n-200)10^{-4}, & n=201,\dots,300, \\
    z_{300}+(n-300)10^{-3}, & n=301,\dots,400.
    \end{array}\r.
\]
For Scho, $N$ is set to 2000, and
\[
z_n=\l\{\begin{array}{ll}
    n\times 10^{-5},          & n=0,   \dots,100,  \\
    z_{100}+(n-100)10^{-4},   & n=101, \dots,1100, \\
    z_{1100}+(n-1100)10^{-2}, & n=1101,\dots,2000.
    \end{array}\r.
\]
Then, the approximation errors between $C^\rho_1$ and (\ref{eq-approxcrho1}) are approximately 6.50$\times10^{-8}$ and 6.07$\times10^{-6}$, respectively, both small enough.

Next, to conduct the MC simulation developed by \cite{AI}, we need to set $M$, the number of time steps, and simulate the value of $S_T$ by dividing the time period $[t,T]$ into $M$ intervals.
More precisely, we denote $\d{t_j:=t+\frac{T-t}{M}j}$ for $j=0,\dots,M$, and first simulate the value of $S_{t_1}$ for given $S_t=\olS$,
and then simulate $S_{t_2}$ using the obtained $S_{t_1}$, repeating this to simulate $S_T$. 
In our numerical experiments, we set $M$ so that $\d{\frac{T-t}{M}=0.01}$.
Besides $M$, we need to set $L$, which is the number of simulation iterations. In our experiments, we set it to 10,000.
Under the above settings, we simulate the value of $F_t(\olS e^{\rho z_n},\olsigma^2+z_n)$ for $n=1,\dots,N$ by using the MC method in \cite{AI}.
In order to approximate the value of the integral $\d{\int_0^\infty F_t(\olS e^{\rho z},\olsigma^2+z)g_\nu(z)dz}$, we compute the following:
\begin{align*}
\lefteqn{F_t(\olS,\olsigma^2)\int_0^{z_1}g_\nu(z)dz} \\
&+\frac{1}{2}\sum_{n=1}^{N-1}(z_{n+1}-z_n)\l[F_t(\olS e^{\rho z_n},\olsigma^2+z_n)g_\nu(z_n)+F_t(\olS e^{\rho z_{n+1}},\olsigma^2+z_{n+1})g_\nu(z_{n+1})\r].
\end{align*}

\begin{figure}[p]
\begin{center}\begin{minipage}{0.8\hsize}\includegraphics[width=100mm]{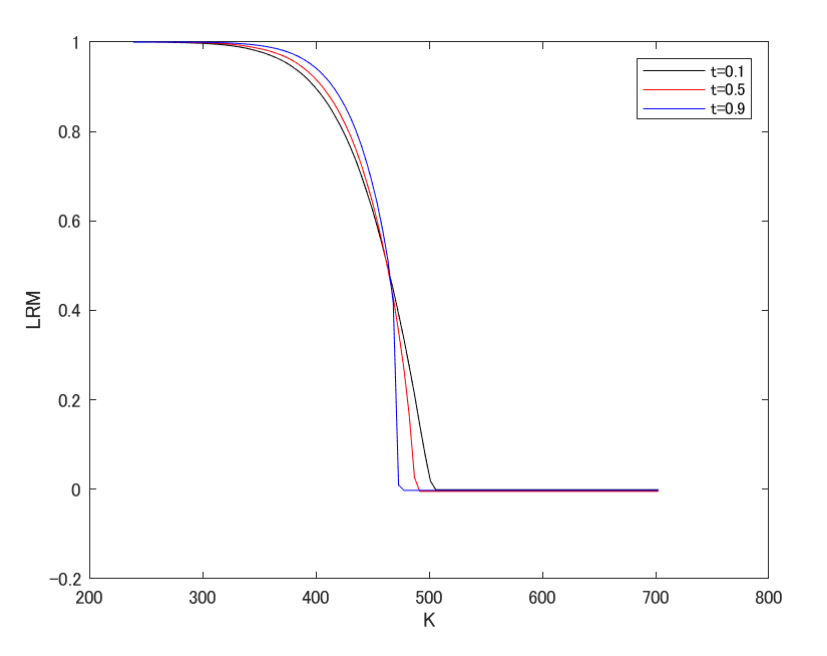}\end{minipage}\end{center}
\begin{center}\begin{minipage}{0.8\hsize}\includegraphics[width=100mm]{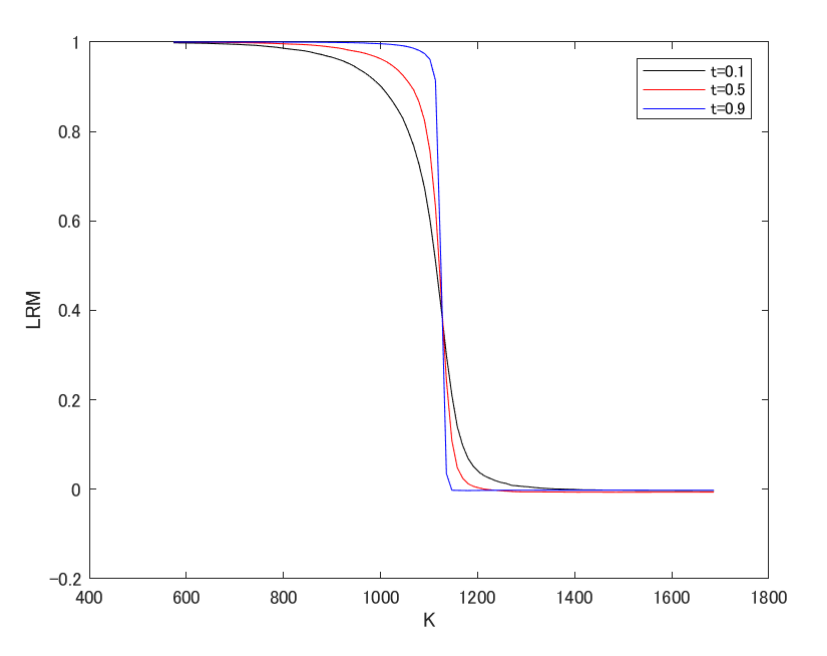}\end{minipage}\end{center}
\caption{The values of LRM strategies for call options vs. strike prices.
The upper and lower panels display the results for the NV and Scho parameter sets, respectively.
The black, red, and blue lines in each panel show LRM strategies at times 0.1, 0.5, and 0.9, respectively.}\label{fig-1}\end{figure}

Here, to make it easier to compare with the results in \cite{AIS-BNS}, we compute LRM strategies for call options instead of put options.
Corollary 3.3 in \cite{AIS-BNS} implies that LRM strategies for call and put options with the same strike price have the following relationship:
\[
\xi_t^{(S_T-K)^+}=1+\xi_t^{(K-S_T)^+}.
\]
For the above two parameter sets, NV and Scho, fixing $T=1$ and varying $K$ from $\dfrac{1}{2}\olS$ to $\dfrac{3}{2}\olS$ with a step of $\dfrac{1}{100}\olS$,
we simulate the values of the LRM strategies $\xi_t^{(S_T-K)^+}$ for a call option with strike price $K$ at times $t=$0.1, 0.5, and 0.9.
We show the results of our numerical experiments in Figure \ref{fig-1}. The obtained results are generally similar to those in Figure 2 of \cite{AIS-BNS}.
In particular, the values of LRM strategies are close to 1 for deep ITM options, 0 for OTM options, and rapidly decreasing around ATM.
Although no way to evaluate the accuracy, these facts indicate that our algorithm is working.
It is worth noting that choosing $N$ and $\{z_n\}_{n=1,\dots,N}$ for which the approximation error to $C^\rho_1$ is not small enough results in poor accuracy, with LRM for OTM options being far from 0.

%%%%%%%%%%%%%%%%%%%%%%%%%%%%%%%%%%%%%%%%%%%%%%%%%%%%%%%%%%%%%%%%%%%%%%%%%%%%%%%

%%%%%%%%%%%%%%%%%%%%%%%%%%%%%%%%%%%%%%%%%%%%%%%%%%%%%%%%%%%%%%%%%%%%%%%%%%%%%%%
%
% Section 5
%
%%%%%%%%%%%%%%%%%%%%%%%%%%%%%%%%%%%%%%%%%%%%%%%%%%%%%%%%%%%%%%%%%%%%%%%%%%%%%%%
\newpage
\setcounter{equation}{0}\section{Concluding remarks}
In this paper, we propose a numerical method for LRM strategies in non-martingale IG-OU type BNS models, using the MC method developed by \cite{AI},
and perform numerical experiments. Still, there are some problems remaining.
The first is that there is no way to evaluate the accuracy of our computation. The second is that we need to restrict the value of $\alpha$ to be near 0.
Indeed, if we set $\alpha=$0.1 in the parameter set NV, the values of LRM strategies for OTM options at $t$=0.1 are approximately 0.3.
Finally, the proposed algorithm is time-consuming. Therefore, it is necessary to make $N$ as small as possible while maintaining accuracy.

%%%%%%%%%%%%%%%%%%%%%%%%%%%%%%%%%%%%%%%%%%%%%%%%%%%%%%%%%%%%%%%%%%%%%%%%%%%%%%%
%                                                                             %
% Appendix A                                                                  %
%                                                                             %
%%%%%%%%%%%%%%%%%%%%%%%%%%%%%%%%%%%%%%%%%%%%%%%%%%%%%%%%%%%%%%%%%%%%%%%%%%%%%%%
\setcounter{section}{0}\renewcommand{\thesection}{\Alph{section}}
\setcounter{equation}{0}\section{Malliavin calculus}\renewcommand{\theequation}{A.\arabic{equation}}
%%%%%%%%%%%%%%%%%%%%%%%%%%%%%%%%%%%%%%%%%%%%%%%%%%%%%%%%%%%%%%%%%%%%%%%%%%%%%%%
Here, we define the Malliavin derivative operator $D_{t,z}$ and present, without proofs, some relevant results that appear in Sections 2 and 3.
In this paper, we use the Malliavin calculus based on the canonical L\'evy space framework by Sol\'e, Utzet, and Vives \cite{S07}.

Suppose that the underlying probability space $(\Omega, \calF, \bbP)$ is given by
\begin{equation}\label{eq-A1}
(\Omega_W\times\Omega_J, \calF_W\times\calF_J, \bbP_W\times\bbP_J),
\end{equation}
where $(\Omega_W, \calF_W, \bbP_W)$ is a one-dimensional Wiener space on $[0,T]$ with coordinate mapping process $W$;
and $(\Omega_J, \calF_J, \bbP_J)$ is the canonical L\'evy space for a subordinator $J$, that is,
\[
\Omega_J=\bigcup_{n=0}^\infty([0,T]\times(0,\infty))^n,
\]
and
\[
J_t(\omega_J)=\sum_{i=1}^nz_i{\bf 1}_{\{t_i\leq t\}}
\]
for $t\in[0,T]$ and $\omega_J=((t_1,z_1),\dots,(t_n,z_n)) \in([0,T]\times(0,\infty))^n$ with the convention that $([0,T]\times(0,\infty))^0$ represents an empty sequence.
Let $\bbF=\{\calF_t\}_{t\in[0,T]}$ be the canonical filtration completed for $\bbP$.

Next, to define the Malliavin derivative operator, we prepare the two measures $q$ and $Q$ defined on $[0,T]\times[0,\infty)$ as follows:
\[
q(E):=\int_E\delta_0(dz)dt+\int_Ez^2\nu(dz)dt,
\]
and
\[
Q(E):=\int_E\delta_0(dz)dW_t+\int_Ez\tN(dt,dz),
\]
where $E\in\calB([0,T]\times[0,\infty))$ and $\delta_0$ is the Dirac measure at $0$.
For $n\in\bbN$, $L_{T,q,n}^2$ denotes the set of product measurable, deterministic functions $h:([0,T]\times[0,\infty))^n\to\bbR$ satisfying
\[
\|h\|_{ L_{T,q,n}^2}^2 := \int_{([0,T]\times[0,\infty))^n}|h((t_1,z_1),\cdots,(t_n,z_n))|^2\prod_{k=1}^nq(dt_k,dz_k)<\infty
\]
with $L_{T,q,0}^2:=\bbR$, and we define
\[
I_n(h):=\int_{([0, T]\times[0,\infty))^n}h((t_1,z_1),\cdots,(t_n,z_n))\prod_{k=1}^nQ(dt_k,dz_k)
\]
for $n\in\bbN$ and $h\in L_{T,q,n}^2$, and $I_0(h):=h$ for $h\in\bbR$. Then, any $X\in L^2(\bbP)$ has the unique representation
$\d{X=\sum_{n=0}^{\infty}I_n(h_n)}$ with functions $h_n\in L_{T,q,n}^2$ that are symmetric in the $n$ pairs $(t_i,z_i), 1\leq i\leq n$,
and we have $\d{\bbE[X^2]=\sum_{n=0}^\infty n!\|h_n\|_{L_{T,q,n}^2}^2}$. Now, we define the Malliavin derivative operator $D_{t,z}$.

%%%%%%%%%%%%%%%%%%%%%%%%%%%%%%%%%%%%%%%%%%%%%%%%%%%%%%%%%%%%%%%%%%%%%%%%%%%%%%%
\begin{defn}
\begin{enumerate}
\item Let $\bbD^{1,2}$ denote the set of $\calF$-measurable random variables $X\in L^2(\bbP)$ with $\d{X=\sum_{n=0}^\infty I_n(h_n)}$ satisfying
      $\d{\sum_{n=1}^\infty nn!\|h_n\|_{L_{T,q,n}^2}^2<\infty}$.
\item For any $X\in\bbD^{1,2}$, the Malliavin derivative $DX:[0,T]\times[0,\infty)\times\Omega\to\bbR$ is defined as
      \[
      D_{t,z}X=\sum_{n=1}^\infty nI_{n-1}(h_n((t,z),\cdot))
      \]
      for $q$-a.e. $(t,z)\in[0,T]\times[0,\infty)$, $\bbP$-a.s.
\end{enumerate}
\end{defn}

Now, we enumerate results related to Malliavin derivatives used in Section 3 without proofs.

%%%%%%%%%%%%%%%%%%%%%%%%%%%%%%%%%%%%%%%%%%%%%%%%%%%%%%%%%%%%%%%%%%%%%%%%%%%%%%%
\begin{prop}[Proposition A.6 of \cite{AIS-BNS}]\label{prop-LT-1}
$\log S_T\in\bbD^{1,2}$ and, we have
\[
D_{t,z}\log S_T=\l\{-\frac{1}{2}\frac{1-e^{-\lambda(T-t)}}{\lambda}+\int_t^T\frac{\sqrt{\sigma_s^2+ze^{-\lambda(s-t)}}-\sigma_s}{z}dW_s+\rho\r\}
\]
for $t\in[0,T]$ and $z>0$.
\end{prop}

%%%%%%%%%%%%%%%%%%%%%%%%%%%%%%%%%%%%%%%%%%%%%%%%%%%%%%%%%%%%%%%%%%%%%%%%%%%%%%%
\begin{lem}[Lemma A.8 of \cite{AIS-BNS}]\label{lem-ut-2}
For any $s\in[0,T]$, we have $u_s\in\bbD^{1,2}$ and
\[
D_{t,z}u_s= \frac{f_u\l(\sqrt{\sigma^2_s+ze^{-\lambda(s-t)}}\r)-f_u(\sigma_s)}{z}{\bf 1}_{[0,s]}(t)
\]
for $t\in[0,T]$ and $z>0$, where $\d{f_u(r):=\frac{\alpha r}{r^2+C_\rho}}$ for $r\in\bbR$.
\end{lem}

%%%%%%%%%%%%%%%%%%%%%%%%%%%%%%%%%%%%%%%%%%%%%%%%%%%%%%%%%%%%%%%%%%%%%%%%%%%%%%%
\begin{lem}[Lemma A.9 of \cite{AIS-BNS}]\label{lem-ut-3}
For any $(s,x)\in[0,T]\times(0,\infty)$, we have $\theta_{s,x}\in\bbD^{1,2}$ and
\[
D_{t,z}\theta_{s,x}= \frac{f_\theta\l(\sqrt{\sigma^2_s+ze^{-\lambda(s-t)}}\r)-f_\theta(\sigma_s)}{z}(e^{\rho x}-1){\bf 1}_{[0,s]}(t)
\]
for $t\in[0,T]$ and $z>0$, where $\d{f_\theta(r):=\frac{\alpha}{r^2+C_\rho}}$ for $r\in\bbR$.
\end{lem}

%%%%%%%%%%%%%%%%%%%%%%%%%%%%%%%%%%%%%%%%%%%%%%%%%%%%%%%%%%%%%%%%%%%%%%%%%%%%%%%
\begin{lem}[Lemma A.10 of \cite{AIS-BNS}]\label{lem-ut-4}
For any $(s,x)\in[0,T]\times(0,\infty)$, we have $\log(1-\theta_{s,x})\in \bbD^{1,2}$ and
\[
D_{t,z}\log(1-\theta_{s,x})=\frac{\log(1-\theta_{s,x}-zD_{t,z}\theta_{s,x})-\log(1-\theta_{s,x})}{z}
\]
for $t\in[0,T]$ and $z>0$.
\end{lem}

%%%%%%%%%%%%%%%%%%%%%%%%%%%%%%%%%%%%%%%%%%%%%%%%%%%%%%%%%%%%%%%%%%%%%%%%%%%%%%%
\begin{prop}[Proposition A.11 of \cite{AIS-BNS}]\label{prop-logZT}
We have $\log Z_T\in\bbD^{1,2}$ and
\begin{align*}
D_{t,z}\log Z_T
&= -\int_t^TD_{t,z}u_sdW_s-\int_t^Tu_sD_{t,z}u_sds-\frac{z}{2}\int_t^T(D_{t,z}u_s)^2ds \\
&  \hspace{5mm}+\int_t^T\int_0^\infty D_{t,z}\log(1-\theta_{s,x})\tN(ds,dx) \\
&  \hspace{5mm}+\int_t^T\int_0^\infty\l(D_{t,z}\log(1-\theta_{s,x})+D_{t,z}\theta_{s,x}\r)\nu(dx)ds+\frac{\log(1-\theta_{t,z})}{z}
\end{align*}
for $t\in[0,T]$ and $z>0$.
\end{prop}

%%%%%%%%%%%%%%%%%%%%%%%%%%%%%%%%%%%%%%%%%%%%%%%%%%%%%%%%%%%%%%%%%%%%%%%%%%%%%%%
\begin{prop}[Proposition 4.1 of \cite{AIS-BNS}]\label{prop-put-deriv}
For any $K>0$, we have $(K-S_T)^+\in\bbD^{1,2}$ and
\[
D_{t,z}(K-S_T)^+=\frac{(K-S_Te^{zD_{t,z}\log S_T})^+-(K-S_T)^+}{z}
\]
for $t\in[0,T]$ and $z>0$.
\end{prop}

%%%%%%%%%%%%%%%%%%%%%%%%%%%%%%%%%%%%%%%%%%%%%%%%%%%%%%%%%%%%%%%%%%%%%%%%%%%%%%%
%\begin{center}
%{\bf Acknowledgments}
%\end{center}
%Takuji Arai and Yuto Imai gratefully acknowledge the financial support of the MEXT Grant-in-Aid for
%Scientific Research (C) No.22K03419 and Early-Career Scientists No. 21K13327, respectively.

%%%%%%%%%%%%%%%%%%%%%%%%%%%%%%%%%%%%%%%%%%%%%%%%%%%%%%%%%%%%%%%%%%%%%%%%%%%%%%%


\begin{thebibliography}{9999}
\bibitem{AI} Arai, T. \& Imai, Y. (2024) Monte Carlo simulation for Barndorff-Nielsen and Shephard model under change of measure, Mathematics and Computers in Simulation, 218, pp.223-234. 
\bibitem{AIS-BNS} Arai, T. \& Imai, Y. \& Suzuki, R. (2017). Local risk-minimization for Barndorff-Nielsen and Shephard models, Finance \& Stochastics, 21, pp.551-592.
\bibitem{AS} Arai, T., Suzuki, R. (2015) Local risk minimization for L\'evy markets. International Journal of Financial Engineering, 2, 1550015.
\bibitem{BNS1} Barndorff-Nielsen, O. E., \& Shephard, N. (2001). Modelling by L\'evy processes for financial econometrics. In L\'evy processes (pp.283-318). Birkh\"auser, Boston, MA.
\bibitem{BNS2} Barndorff-Nielsen, O. E., \& Shephard, N. (2001). Non-Gaussian Ornstein-Uhlenbeck-based models and some of their uses in financial economics.
               Journal of the Royal Statistical Society: Series B (Statistical Methodology), 63(2), pp.167-241.
\bibitem{NV} Nicolato, E. \& Venardos, E. (2003). Option pricing in stochastic volatility models of the Ornstein-\"Uhlenbeck type, Mathematical Finance, 13, pp.445-466.
\bibitem{SP} Sabino, P., \& Petroni, N. C. (2022). Fast simulation of tempered stable Ornstein-\"Uhlenbeck processes. Computational Statistics, 37(5), pp.2517-2551.
\bibitem{Scho} Schoutens, W. (2003). L\'evy processes in finance: pricing financial derivatives, Wiley.
\bibitem{Sch3} Schweizer, M. (2008). Local Risk-Minimization for Multidimensional Assets and Payment Streams. Banach Center Publ., 83, pp.213-229.
\bibitem{S07} Sol\'e, J.L., Utzet, F., \& Vives, J. (2007). Canonical L\'evy process and Malliavin calculus. Stochastic Process. Appl., 117, pp.165-187.
\end{thebibliography}
\end{document}